\documentclass[twocolumn,aps, prd, amsmath, floats,floatfix, superscriptaddress, nofootinbib, showkeys,
showpacs]{revtex4}
\usepackage{amssymb}
\usepackage{amsmath}
\usepackage{verbatim}
\usepackage{mathrsfs}
\usepackage{amsfonts}
\usepackage{latexsym}
\usepackage{epsfig}
\usepackage{color}
\usepackage{graphicx,subfigure}
\usepackage{units}
\usepackage{envmath}
\usepackage{natbib}
\usepackage{ctable}
\begin{document}
%%%%%%%%%%%%%%%%%%
%%%   MACROS   %%%
%%%%%%%%%%%%%%%%%%
\definecolor{orange}{rgb}{0.9,0.45,0}
\def\CovDev{D}
\def\Res{{\mathcal R}}
\def\Gammaflat{\hat \Gamma}
\def\metricflat{\hat \gamma}
\def\Dflat{\hat {\mathcal D}}
\def\part_n{\partial_\perp}
%
%=== Definition for abbreviations ===
\def\Lie{\mathcal{L}}
\def\A{\mathcal{X}}
\def\Aphi{\A_{\phi}}
\def\hAphi{\hat{\A}_{\phi}}
\def\E{\mathcal{E}}
\def\Ham{\mathcal{H}}
\def\M{\mathcal{M}}
\def\R{\mathcal{R}}
\def\p{\partial}
\def\hg{\hat{\gamma}}
\def\hA{\hat{A}}
\def\hD{\hat{D}}
\def\hE{\hat{E}}
\def\hR{\hat{R}}
\def\hcA{\hat{\mathcal{A}}}
\def\hDelt{\hat{\triangle}}
\def\na{\nabla}
\def\dif{{\rm{d}}}
\def\non{\nonumber}
\newcommand{\erf}{\textrm{erf}}
%====================================
%
\renewcommand{\t}{\times}
\long\def\symbolfootnote[#1]#2{\begingroup%
\def\thefootnote{\fnsymbol{footnote}}\footnote[#1]{#2}\endgroup}
\title{Primordial black hole merger rate in self-interacting dark matter halo models}

\author{Saeed Fakhry} 
\email{s_fakhry@sbu.ac.ir}
\affiliation{Department of Physics, Shahid Beheshti University, Evin, Tehran 19839, Iran}
\author{Mahdi Naseri}
\email{mahdi.naseri@email.kntu.ac.ir}
\affiliation{Department of Physics, K.N. Toosi University of Technology, P.O. Box 15875-4416, Tehran, Iran}
\author{Javad T. Firouzjaee}
\email{firouzjaee@kntu.ac.ir}
\affiliation{Department of Physics, K.N. Toosi University of Technology, P.O. Box 15875-4416, Tehran, Iran}
\affiliation{School of Physics, Institute for Research in Fundamental Sciences (IPM), P.O. Box 19395-5531, Tehran, Iran}
\author{Mehrdad Farhoudi}
\email{m-farhoudi@sbu.ac.ir}
\affiliation{Department of Physics, Shahid Beheshti University, Evin, Tehran 19839, Iran}

\date{\today}
\begin{abstract}
We study the merger rate of primordial black holes (PBHs) in self-interacting dark matter (SIDM) halo models. To explore a numerical description of the density profile of SIDM halo models, we use the result of a previously performed simulation for SIDM halo models with $\sigma/m=10~\rm cm^{2}g^{-1}$. We also propose a concentration-mass-time relation that can explain the evolution of the halo density profile related to SIDM models. Furthermore, we investigate the encounter condition of PBHs that may have been randomly distributed in the medium of dark matter halos. Under these assumptions, we calculate the merger rate of PBHs within each halo considering SIDM halo models and compare the results with that obtained for cold dark matter (CDM) halo models. To do this, we employ the definition of the time after halo virialization as a function of halo mass. Also, by unifying the transition time for all halos using a quantity that depends on the halo mass and cross section of SIDM particles, we indicate that the SIDM halo models, for $f_{\rm PBH}>0.32$, can generate sufficient PBH mergers in a way that those exceed the one resulted from CDM halo models for $f_{\rm PBH}=1$. By considering the spherical-collapse halo mass function, we obtain similar results for the cumulative merger rate of PBHs. Moreover, we calculate the evolution of the PBH total merger rate as a function of redshift. The results show that SIDM models when considering $f_{\rm PBH}>0.32$, could have generated more significant PBH mergers than CDM models at higher redshifts. To determine a constraint on the PBH abundance, we study the merger rate of PBHs in terms of their fraction and masses and compare those with the black hole merger rate estimated by the Advanced Laser Interferometer Gravitational-Wave Observatory (aLIGO)-Advanced Virgo (aVirgo) detectors during the third observing run. The results demonstrate that within the context of SIDM halo models, the merger rate of $10\,M_{\odot}\mbox{-}10\,M_{\odot}$ events can potentially fall within the aLIGO-aVirgo window. We also estimate a relation between the fraction of PBHs and their masses, which is well consistent with our findings. 
\end{abstract}


\maketitle

\vspace{8cm}

%\tableofcontents

\section{Introduction} \label{sec.i}

Over the past few years, the Laser Interferometer Gravitational-Wave Observatory (LIGO)-Virgo Collaboration has detected gravitational waves emitted by about $50$ inspiraling and merging black hole binaries \cite{Abbott:2016blz, Abbott:2016nmj, TheLIGOScientific:2016src,TheLIGOScientific:2017qsa, Abbott:2020khf, Abbott:2020tfl}, which has opened a new epoch in probing the nature and behavior of compact objects in the Universe. Interestingly, most of the black hole mergers recorded by the LIGO-Virgo detectors are related to black holes with masses around $30\,M_{\odot}$. This fact certainly provides suggestive information on the mass distribution of black holes in the Universe.

Still, we do not know much about the origin of these black holes. There is a possibility that they are ordinary astrophysical black holes from stellar collapses (possibly from different channels) \cite{Rodriguez:2021nwd,Fishbach:2021yvy}. The other interesting conjecture is that the LIGO-Virgo detectors have detected primordial black holes (PBHs). These gravitational-wave observatories are continuing to probe the population of black holes, seeking to specify whether the mergers provide any direct evidence for the existence of PBHs.

PBHs could have formed in an early period of the evolution of the Universe as a consequence of the gravitational collapse of cosmological perturbations \cite{Carr:1975qj,Zeldovich:1967lct}. Under such a hypothesis, PBHs can be generated due to high nonlinear rare peaks in the primordial distribution of density perturbations produced during the inflationary era. These perturbations can finally collapse when reentering the horizon, and produce black holes during the radiation-dominated era and/or some transitional matter phase. To form PBHs, these cosmological perturbations need to have some critical states. Passing from the threshold value of the density is the critical state of formation. Many numerical investigations have been performed to study the threshold value for the density perturbations, see, e.g., Refs.~\cite{Carr:1975qj, Niemeyer:1999ak,Young:2014ana, Young:2, Shibata:1999zs, musco, bloom,allah}.

Before the detection of gravitational waves, many studies were done on the subject of PBHs as a candidate for dark matter; see, e.g., Ref.~\cite{Carr:2020xqk} and references therein. This issue comes from the fact that massive PBHs interact only via gravitation, and since a large number of black holes have fluid behavior on sufficiently large scales, PBHs are a natural candidate for dark matter. However, nowadays, the very strong observational limits on the abundance of PBHs are themselves a powerful and unique method of investigating the early Universe at small scales, which cannot be tested by any other method~\cite{Lehmann:2018ejc, Carr:2017jsz, Carr:2020gox}. Nevertheless, the existence of PBHs has been neither proven nor refuted.

Assuming involving PBHs in merging pairs, serious bounds on the abundance of PBHs in the mass range around $10\mbox{-}30\,M_{\odot}$ can be obtained from the LIGO-Virgo observations. Shortly after the first observation of a binary black hole merger, several groups of researchers claimed that the merger rate obtained from the LIGO-Virgo discovery is potentially consistent with a mass fraction of PBHs accounting for the total of dark matter~\cite{bird, Clesse:2016vqa}. The main assumption of their studies was that any two involved black holes had a primordial origin and the LIGO-Virgo detectors had detected dark matter. Assuming that PBHs are a fraction of dark matter and their merging happens in the dark matter halo, the halo mass function can affect the merger rate of PBHs \cite{Fakhry:2020plg}. Also, any change in the concentration parameter can have an effect on the relative velocity distribution of PBHs within each halo, which determines the PBH merger rate within each halo. Accordingly, one can expect that different dark matter halo models will have different predictions for the PBH merger rates.

One of the most renowned dark matter halo models is the self-interacting dark matter (SIDM) halo model, which can resolve many astrophysical problems \cite{Spergel:1999mh} and has rapidly turned into an interesting alternative to the cold dark matter (CDM) halo model. One famous example of these types of problems is the ``core-cusp problem", which stems from a discrepancy between observations and CDM simulations of the halo density profiles. That is, the CDM simulations indicate a steep slope (cusp) for the density profile in the central regions of halos, however, the observed rotational curves of stars in galaxies reveal a very low slope (core) for the density profile. 

The ``missing satellite problem" is another challenge for the CDM paradigm. CDM simulations predict the existence of a myriad of subhalos around the Milky Way (approximately $500$ dwarf spheroidal galaxies \cite{Moore:1999nt}), yet the current number of the observed dwarf spheroidal galaxies is only approximately $50$ \cite{Bullock}. These problems can be solved by considering a collisional type of dark matter with a non-negligible cross section per unit mass of particles instead of collisionless CDM. Since dark matter particles have a non-negligible cross section in SIDM halos, they can interact with themselves. As a result, one can see the dependence of the halo density profile on the dark matter cross section~\cite{Bernal:2020kse}. Indeed, it has been established \cite{Spergel:1999mh} that such an assumption may result in a heat transfer, which reduces the density of the central regions of halos and consequently eliminates the core-cusp problem. The capability of interaction with the other dark matter particles inside halos also leads to the formation of fewer satellites, and removes the missing satellite problem.

Considering hard-sphere scattering as the type of collision among the dark matter particles, the primary role in determining the SIDM properties is played by the constant parameter of $\sigma/m$, that is the value of cross section per unit mass of dark matter particles. Alternatively, $\sigma/m$ can depend on the velocity and interact via a Yukawa potential \cite{Loeb:2010gj}. The idea of velocity-dependent models of SIDM stems from some constraints on $\sigma/m$, which state that velocity-independent models of SIDM fail to solve small-scale astrophysical problems \cite{Miralda-Escude:2000tvu,Gnedin:2000ea}. However, even if velocity-independent models are unable to resolve astrophysical problems, a velocity-dependent model based on a Yukawa potential may still behave similarly to a velocity-independent model in some range given particular assumptions. Considering the mediator mass $m_\phi$ to be greater than one percent of the dark matter particle mass, the scattering cross section becomes independent of velocity for almost all astrophysically relevant velocities of dark matter \cite{Robertson:2018anx}. The collisional dark matter that is studied in this work has a fixed $\sigma/m$ and does not change with the velocity and we neglect the open discussions on the dependence of the dark matter cross section on its velocity and its compatibility with various constraints.

It is known that dark matter constitutes a remarkable portion of the mass/energy of the Universe. Thus, it is plausible to think of it as being made of more than just one component, such as both PBHs and SIDM. It has been demonstrated that even if a small portion of dark matter is strongly self-interacting, e.g., if only less than $10\%$ of dark matter has $\sigma/m>1~\rm cm^2 g^{-1}$ and the rest is not self-interacting, the model can remove astrophysical problems and still keep the benefits of a completely SIDM model \cite{Pollack:2014rja}. In other words, it is also possible to consider a sector of dark matter as strongly self-interacting in addition to the rest of it being made of PBHs and even other components, avoiding violating the astrophysical constraints.

In this work, we propose to use SIDM halo models to calculate the merger rate of PBHs. In this respect, the outline of the work is as follows. In Sec.~\ref{sec:ii} we propose a halo model for the SIDM scenario, which includes a convenient density profile, concentration-mass-time relation, and the spherical-collapse halo mass function. Then, in Sec.~\ref{sec:iii} we calculate the merger rate of PBHs in SIDM halo models, and compare it with the corresponding results of CDM halo models. We also explore the redshift evolution of the PBH merger rate for the SIDM halo models and compare it with the findings from CDM halo models. Furthermore, we present constraints on the fraction of PBHs arising from SIDM models. Finally, we discuss the results and summarize the findings in Sec.~\ref{sec:iv}.

%%%%%%%%%%%%%%%%%%%%%%%%%%%%%%%%%%%%%%%%%%%%%%

\section{Halo models}
\label{sec:ii}
\subsection{Halo density profile}\label{sec.iia}
For CDM halo models, many simulations have been performed and resulted in some functions for their density profile. One of the most famous and successful density profiles was proposed by Navarro, Frenk, and White (NFW) \cite{Navarro:1996gj}, which has the form
\begin{equation} \label{NFW}
	\frac{\rho}{\rho_{\rm 0}}=\frac{1}{x(1+x)^2} \, ,
\end{equation}
where $x\equiv r/r_{\rm s}$ and $r_{\rm s}$ is the scale radius of the halo. In addition, $\rho_{\rm 0}$ is given by $\rho_{\rm 0}=\rho_{\rm crit}\delta_{\rm c}$ for each halo, where $\rho_{\rm crit}$ is the critical density of the Universe at a given redshift $z$ and $\delta_{\rm c}$ is the linear threshold of overdensities that depends on the concentration parameter $C$ with the relation
\begin{equation} \label{delta_c}
	\delta_{\rm c}=\frac{200}{3}\frac{C^3}{\ln(1+C)-C/(1+C)} \, .
\end{equation}
The concentration parameter is basically defined as the ratio of the virial radius of the halo, $r_{\rm vir}$, to its scale radius, $r_{\rm s}$. On the other hand, $r_{200}$ is defined as the radius of the volume inside which the mean density is roughly equal to 200 times the critical density of the Universe. This radius is usually considered close to the virial radius of the halo, as the overdensity of the halo at the time
of virialization is close to approximately $200$ \cite{Bryan}. Therefore, the concentration parameter is
\begin{equation} \label{r_200}
	C\equiv\frac{r_{\rm vir}}{r_{\rm s}}\simeq\frac{r_{\rm 200}}{r_{\rm s}} \, .
\end{equation}

It can be deduced from Eq.~(\ref{NFW}) that the NFW density profile predicts $\rho \propto r^{-1}$ when $x\to 0$, whereas the density changes with radius as $\rho \propto r^{-3}$ when $x\to \infty$. The SIDM simulations are significantly different than the CDM simulations of the density profile and consequently do not agree with the NFW behavior. One of the SIDM simulations was performed by Fischer {\it et al.} and its outcome was provided in Fig.~5 of Ref.~\cite{Fischer:2020uxh}. This simulation includes the SIDM particles with a cross section per unit mass of $\sigma/m=10~\rm cm^2 g^{-1}$ and was run for halos with mass $10^{15}\,M_{\odot}$. According to its results, the density profile gradually deviates from the NFW one in the inner region of the halo, and becomes cored by $1~\rm Gyr$ after the halo virialization time. After this time, the central density grows with time, due to the fact that self-interactions among dark matter particles lead to energy transfer from the inner region to the outer parts of the halo. The core contracts due to this energy loss, and slowly becomes denser at the center, which is the so-called gravothermal core collapse in the literature {\cite{Fischer:2020uxh}}.

Our first task in this work is to find a quantitative description for the density profile of SIDM halo models for calculating the merger rate of PBHs while assuming that the dark matter particles interact with themselves. In fact, the main effect of SIDM halo models in our study stems from the evolution of the halo density profile. For this purpose, we try to specify a function that fits the result of the mentioned simulation. This issue can be performed by exerting some modifications to the NFW density profile.

The mentioned simulation only consists of one mass window, while we need to study a range of halo masses. Other simulations illustrate that both core formation and core collapse happen more rapidly when the halos are more massive \cite{Robertson:2018anx}. To have an identical time evolution for all halos, it is useful to define the new units
\begin{eqnarray}\label{t_0}
t_{0}^{-1}=\frac{2.26\,\sigma}{\sqrt{2} \pi} \sqrt{\frac{G M_{\rm vir}^{3}}{r_{\rm s}^{7}}}, \hspace*{0.5cm} \rho_{0}=\frac{M_{\rm vir}}{2 \pi r_{\rm s}^{3}}.
\end{eqnarray}
where $\sigma$ is the cross section of SIDM particles, $G$ is the gravitational constant, and $M_{\rm vir}$ is the virial mass of halos. Then, if density and time are expressed in these units, the evolution of the density profile remains the same for all halos with various masses \cite{Vogelsberger}, as $t_{0}$ and $\rho_{0}$ depend on halo mass (see the top panel of Fig.~2 in Ref.~\cite{Vogelsberger}).

\begin{figure}[t!]
	\begin{minipage}{1\linewidth}
		\includegraphics[width=1\textwidth]{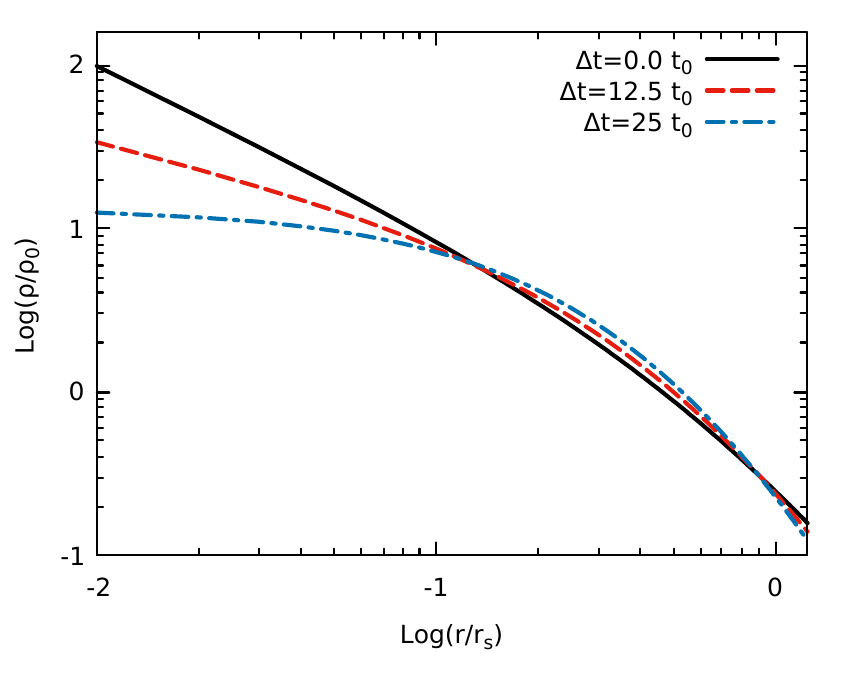}
	\\ \hspace*{0.5cm} \\ 
		\includegraphics[width=1\textwidth]{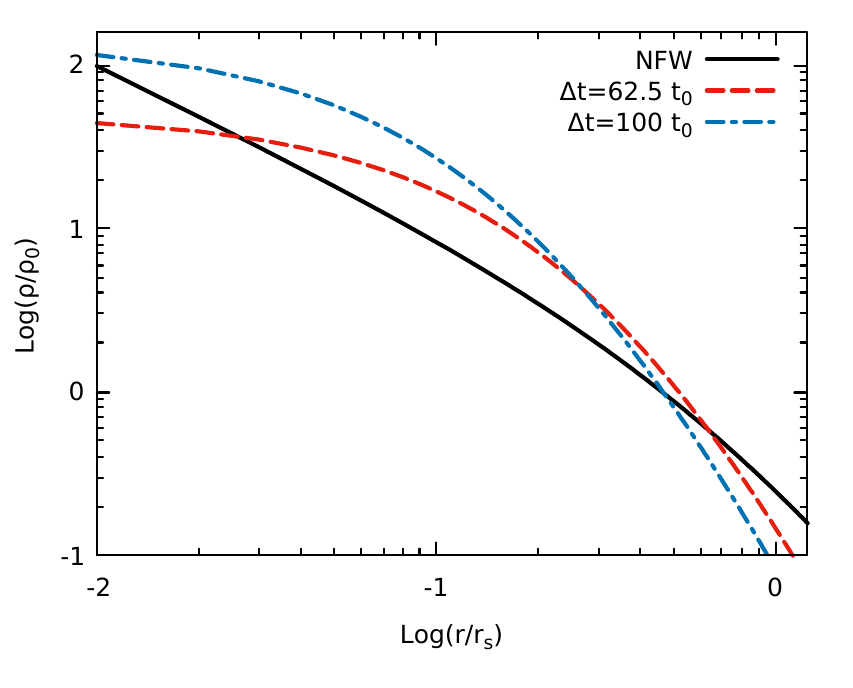}
		\caption{Density profiles of Eqs.~(\ref{early_time}) and (\ref{late_time}) for several times after the halo virialization for SIDM halo models. The top plot reveals how the density profile becomes cored before $\Delta t=25\,t_0$, and the bottom plot shows how it changes after this time. At $\Delta t=0$, the proposed profile overlaps the NFW density profile. The NFW density profile is also displayed in the bottom panel for comparison.}
		\label{figure_1}
	\end{minipage}
\end{figure}

Actually, the SIDM and CDM models are the same at $\Delta t=0$, while at later times the SIDM model deviates from the CDM model. In other words, the process of core formation in the SIDM model starts at $\Delta t > 0$, and the core is formed at $\Delta t = 25\,t_{0}$ after halo virialization~\cite{Vogelsberger}, which is $1$ Gyr for halos with $M_{\rm vir}=10^{15}~M_{\odot}$ and $\sigma/m=10~\rm cm^2 g^{-1}$. Therefore, $25\,t_{0}$ is the transition time between two different epochs. We divide the problem into these two epochs, i.e. before and after $25\,t_{0}$ elapsed from the halo virialization time, and find a properly fitted model that describes the simulations of Refs.~\cite{Fischer:2020uxh,Vogelsberger}. Under these considerations, for the first epoch (i.e., before $25\,t_{0}$), we take the following form of the density profile:
\begin{equation} \label{early_time}
	f(x)=\dfrac{\rho}{\rho_{\rm 0}}=\dfrac{1+G_{11}(1-\alpha_{1})e^{-x/l_{1}}}{x^{\alpha_{1}}(1+G_{12}x)^{3-\alpha_{1}+\beta_{1}}} \, ,
\end{equation}
where $G_{11}$, $G_{12}$, $\alpha_{1}$, $\beta_{1}$, and $l_{1}$ are free parameters to be determined. We have found some numerical suggestions in a way to have three criteria as: $(i)$ The density profile being reduced to the NFW one at $\Delta t=0$; $(ii)$ It forms core at $\Delta t=25\,t_{0}$; $(iii)$ It continuously changes in this interval. Clearly, all of the free parameters are only functions of the time elapsed from the halo virialization. Under these conditions, one can set the free parameters of Eq. (\ref{early_time}) to be
\begin{eqnarray}\label{parameters_1}
	& \alpha_{1} = 1-\dfrac{\Delta t}{25\,t_0} \, , \kern 2.3pc \beta_1 = \dfrac{\Delta t}{25\,t_0} \, , \kern 2.3pc l_{1}=4 \, , \nonumber \\  \nonumber \\ 
	& G_{11}=24\exp\left[-\left(\dfrac{\frac{\Delta t}{25\,t_0} -1.938}{1.148}\right)^2\right] \, ,  \\ \nonumber \\ 
	& G_{12}=0.6 \, \dfrac{\Delta t}{25\,t_0} +1 \, . \nonumber 
\end{eqnarray}
Then, we iterate this method for the second epoch, i.e., $\Delta t>25\,t_0$. Accordingly, the second function must continuously change from the cored shape until forming the third plot in Fig.~5 of Ref.~\cite{Fischer:2020uxh} at $\Delta t=100\,t_0$ ($\Delta t=4~\rm Gyr$ for $M_{\rm vir}=10^{15}\,M_{\odot}$). Furthermore, both functions must reduce to each other at $\Delta t=25\,t_0$ for continuity condition. Hence, under these conditions, the density profile for the second epoch (that is, $\Delta t>25\,t_0$) is
\begin{equation} \label{late_time}
f(x)=\frac{\rho}{\rho_{\rm 0}}=\dfrac{1+G_{21}(1-\alpha_{2})e^{-x/l_{2}}}{x^{\alpha_{2}}(1+G_{22}x)^{3-\alpha_{2}+\beta_{2}}} \, ,
\end{equation}
where $G_{21}$, $G_{22}$, $\alpha_{2}$, $\beta_{2}$, and $l_{2}$ are also free parameters to be determined. Similarly, to specify the best fit of the density profile simulated in Ref.~\cite{Fischer:2020uxh}, we have also suggested the expressions for those free parameters to be
\begin{eqnarray}\label{parameters_2}
	& \alpha_2 = 0 \, ,  \kern 2.3pc \beta_2 = 1 \, , \kern 2.3pc l_{2}=4 \, , \nonumber \\ \nonumber \\ 
	& G_{21}=12.3 \, \left(\dfrac{\Delta t}{25 \, t_0}\right)^{1.75} \, , \\ \nonumber \\ 
	& G_{22}=11.38 \, \exp\left[-\left(\dfrac{\frac{\Delta t}{25 \, t_0} -9.544}{6.11}\right)^2\right] \, . \nonumber
\end{eqnarray}
The resulted density profiles, i.e., Eqs.~(\ref{early_time}) and (\ref{late_time}), are shown in Fig.~\ref{figure_1} for different $\Delta t$ values. Comparing this figure with the result obtained in Ref.~\cite{Fischer:2020uxh} shows that our suggested density profile and its parameters are in good agreement with the mentioned simulation. It should be noted that the obtained density profile is consistent with the NFW profile in the noninteracting limit of dark matter. This limit, that is, $\sigma/m \to 0$, is equivalent to $t_0 \to \infty$ given Eq.~(\ref{t_0}). For a very large $t_0$, Eq.~(\ref{early_time}) remains as the density profile of halo for a very long time (i.e., $25\,t_0 \to \infty$). One can easily see that $t_0 \to \infty$ leads to $\alpha_1 \to 1$, $\beta_1 \to 0$, and $G_{12} \to 1$, whereby this equation converges to the NFW profile.

To find more exact expressions for the parameters of the proposed density profiles, the repetition of the simulation for a number of $\Delta t$ values is required. Moreover, the simulation is conducted for SIDM particles with cross section per unit mass of $\sigma/m=10~\rm cm^2 g^{-1}$. This means that the proposed density profiles are only acceptable for this value. However, many other simulations predict that lower values of $\sigma/m$ merely affect the speed of core formation and core collapse \cite{Brinckmann:2017uve}. In other words, the shapes of the profiles do not change as every stage occurs at a later time compared with $\sigma/m=10~\rm cm^2 g^{-1}$. However, there is an exception, i.e., the effects of SIDM interactions are negligible for a lower limit of the cross section per unit mass of the dark matter particle. Thus, for very small values of $\sigma/m$, deviation from the NFW profile is not considerable. Clearly, in the purposed studying of the impact of the SIDM on any astrophysical phenomenon, the dark matter particles are considered to have larger cross section than the lower limit (which has different values in various studies), otherwise, there would be no observable difference between the CDM and SIDM halo models.
%%%%%%%%%%%%%%%%%%%%%%%%%%%%%%%%%%%%%%%%%%%%%%%%%%%%%%%%
\subsection{Halo concentration-mass-time relation}\label{subsec:HMCT}
The primary conclusion of studying the density profiles of SIDM halo models reveals that, while the density profile remains unchanged with time in CDM halo models, it dramatically changes with time for SIDM halo models. Such a difference means that the concentration parameter is no longer time independent and changes with time for SIDM halos.

In addition, $N$-body simulations indicate that the concentration parameter is a decreasing function of the halo mass~\cite{prada, Dutton:2014xda, Okoli:2015dta, Ludlow:2016ifl}. For CDM halo models, a relation has been obtained between the concentration parameter and the halo mass \cite{Maccio:2008pcd},
\begin{equation} \label{m_c_NFW}
	C_{\rm NFW}=8.3\left(\frac{M_{200}}{10^{12}h^{-1}M_{\odot}}\right)^{-0.104} \, .
\end{equation}
Here, $M_{200}$ is the mass enclosed by the radius $r_{200}$, which is usually close to the virial mass, $M_{\rm vir}$, at the halo virialization time~\cite{Bryan}.

It should be noted that the scale radius of a halo, $r_{\rm s}$, is defined as the radius at which the logarithmic slope of the density profile is $-2$, i.e., $\rho(x_{\rm s}) \propto x^{-2}$~\cite{Bondarenko:2017rfu}. In order to determine the scale radius in the proposed density profiles, the method provided in Ref.~\cite{Naseri:2020dgq} can be used. Hence, we demand that the logarithmic slope of the density distribution is $-2$ at the scale radius, that is,
\begin{equation} \label{x_s}
	\frac{d\ln\rho(x)}{d\ln(x)}\vert_{x=x_{\rm s}}=\frac{d\ln(x^{-2})}{d\ln(x)}\vert_{x=1}=-2 \, .
\end{equation}
Accordingly, $x_{\rm s}$ can be obtained via numerical solution. As the density profile for both models (i.e., CDM and SIDM) is the same at the halo virialization time (i.e., at $\Delta t=0$), $r_{\rm vir}$ is the same for both cases and $C$ changes with $r_{\rm s}$ regarding Eq. (\ref{r_200}). Therefore,
\begin{equation} \label{c_new1}
	\frac{C_{\rm NFW}}{C}=\frac{r_{\rm s}}{r_{\rm s,NFW}}=x_{\rm s} \, .
\end{equation} 
By specifying the scale radius via Eq.~(\ref{x_s}), Eq.~(\ref{c_new1}) can be used together with Eq.~(\ref{m_c_NFW}) to calculate the halo concentration parameter for the given virialized mass $M_{\rm vir}$ and for a time $\Delta t$ after the halo virialization.

\begin{figure}[t!]
	\begin{minipage}{1\linewidth}
		\includegraphics[width=1.0\textwidth]{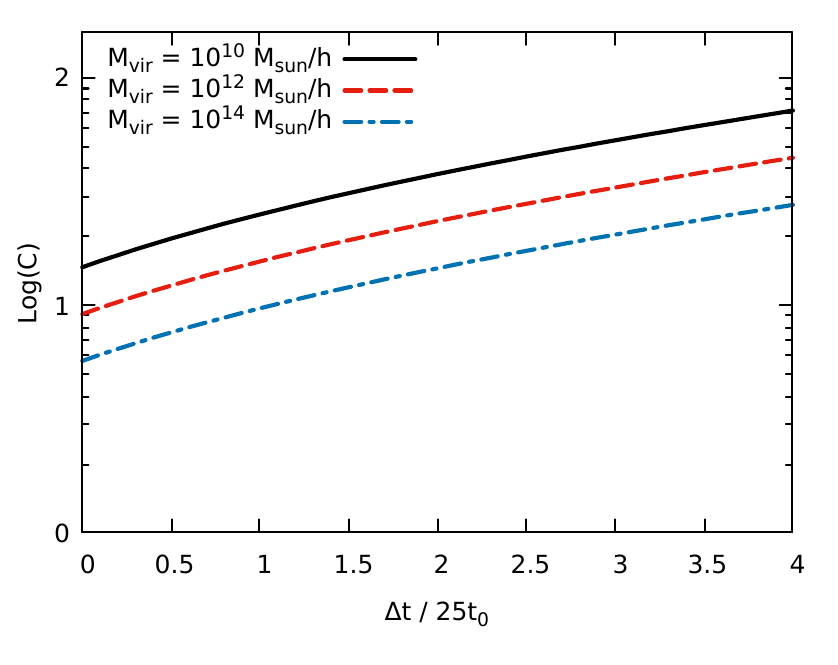}
		\caption{Evolution of the concentration parameter with respect to the time after halo virialization. The solid (black), dashed (red), and dot-dashed (blue) lines show this relation for the virial masses of $M_{\rm vir}=10^{10}$, $10^{12}$, and $10^{14}\,h^{-1}M_{\odot}$, respectively.}
		\label{figure_2}
	\end{minipage}
\end{figure}

\begin{figure}[t!]
	\begin{minipage}{1\linewidth}
		\includegraphics[width=1\textwidth]{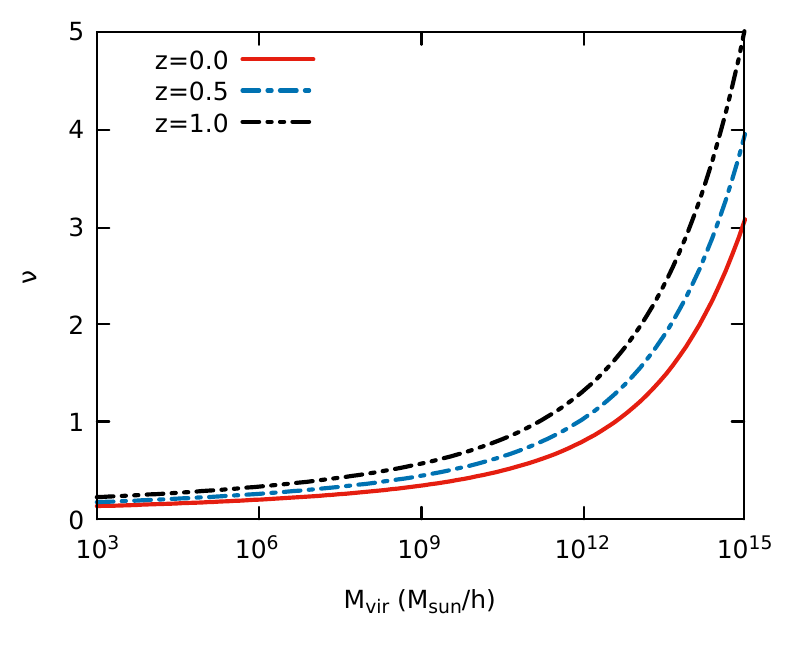}
		\caption{Peak height as a function of halo mass and redshift. The solid (red), dot-dashed (blue), and dot-dot-dashed (black) lines represent the $\nu(M)$ relation for $z=0,~0.5$, and $1$, respectively.}
		\label{fig:nu-m}
	\end{minipage}
\end{figure}

We use the above method for several halo mass values and a range of times after the halo virialization, to calculate the concentration parameter, and then try to specify the best fit of $C(\Delta t/25\,t_0, M_{\rm vir})$ for the concentration-mass-time relation. The following function properly fits the result and can be considered as the concentration-mass-time relation for SIDM halo models:
\begin{equation}\label{M_c_t}
C_{\rm SIDM}= \{k_1 \exp(k_2 \, \frac{\Delta t}{25 \, t_0}) + k_3\}\left(\frac{M_{\rm vir}}{10^{14}h^{-1}M_{\odot}}\right)^{k_4} \, ,
\end{equation}
with $k_1 = 18.53$, $k_2 = 0.1951$, $k_3 = -12.84$, and $k_4 = - 0.104$. The result is shown in Fig.~\ref{figure_2} for three different halo masses. In agreement with many other studies \cite{Bondarenko:2017rfu, Sagunski:2020spe}, SIDM halo models lead to a more significant concentration parameter than the corresponding one obtained from CDM halo models.
Moreover, in order to calculate the redshift evolution of the concentration-mass relation, we refer to the definition of the linear root-mean-square fluctuation of overdensities,
\begin{equation}\label{sigma}
	\sigma^{2}(M,z) \equiv \frac{1}{2\pi^{2}}\int_{0}^{\infty}P(k,z)W^{2}(k,M)k^{2}dk,
\end{equation}
where $P(k,z)$ is the power spectrum of the fluctuations, and $W(k, M)$ is the Fourier spectrum of the top-hat filter that depends on the halo mass $M$ and the wave number $k$. Using this description, one can define a dimensionless parameter called the peak height, $\nu(M, z)$, as
\begin{equation}\label{nu}
	\nu(M, z) \equiv \frac{\delta_c}{\sigma(M, z)}=\frac{\delta_c}{D(z)\sigma(M,0)},
\end{equation}
where $\delta_{\rm c}=1.686$ is the linear threshold of overdensities for the spherical-collapse halo models. Also, $D(z)$ is the linear growth factor \cite{Lahav:1991wc}, given by
\begin{equation}\label{growth}
D(z)=\frac{\Omega_{m}(z)}{\Omega_{m}(0)}\frac{\psi(0)}{\psi(z)}(1+z)^{-1},
\end{equation}
where $\psi(z)$ is defined as
\begin{equation}
\psi(z)=\Omega_m(z)^{4/7}-\Omega_{\Lambda}(z)+\left(1+\frac{\Omega_{m}(z)}{2}\right)\left(1+\frac{\Omega_{\Lambda}(z)}{70}\right).
\end{equation}
In the above relations, $\Omega_{\Lambda}(z)$ and $\Omega_m(z)$ are dark energy and matter density parameters, which are defined as
\begin{equation}
\Omega_{\Lambda}(z)=\frac{\Omega_{\Lambda}(0)}{\Omega_{\Lambda}(0)+\Omega_{m}(0)(1+z)^{3}}, \hspace*{0.5cm} \Omega_{m}(z)=1-\Omega_{\Lambda}(z).
\end{equation} 
Specifically, the peak height depends on the halo mass and redshift. On the other hand, for halo masses within the range $10^{-7}\leqslant M_{\rm vir}/(h^{-1}M_{\odot}) \leqslant10^{15}$, the linear root-mean-square fluctuation of overdensities can be approximated as \cite{Ludlow:2016ifl}
\begin{equation}\label{sigma_m_vir}
	\sigma (M, z) \simeq D(z)\frac{22.26 \chi^{0.292}}{1+1.53 \chi^{0.275} + 3.36 \chi^{0.198}} \, ,
\end{equation}
where
\begin{equation}\label{m_vir}
	\chi = \left(\frac{M_{\rm vir}}{10^{10} h^{-1}M_{\odot}}\right)^{-1} \,.
\end{equation}
Using Eqs~(\ref{nu}), (\ref{sigma_m_vir}), and (\ref{m_vir}), one can calculate the peak height parameter in terms of the halo mass and the linear growth factor to be
\begin{equation} \label{c-nu}
	\nu(M_{\rm vir}, z) \simeq \frac{1}{D(z)}\sum_{i=1}^{3} a_{i} \chi ^{b_{i}}\, ,
\end{equation}
where $a_{1}=0.0757$, $a_{2}=0.254$, $a_{3}=0.115$, $b_{1}=-0.292$, $b_{2}=-0.094$ and $b_{3}=-0.017$. Fig.~\ref{fig:nu-m} shows the peak height as a function of the halo mass and redshift. Eventually, using Eq.~(\ref{c-nu}), one can obtain the $C(\nu)$ relation, which enables us to provide the concentration parameter as a function of the halo mass and redshift. In particular, we will use this relation to determine the redshift evolution of the total merger rate of PBHs (see Sec.~\ref{sec:iiib}).

\begin{figure}[t!]
	\begin{minipage}{1\linewidth}
		\includegraphics[width=1\textwidth]{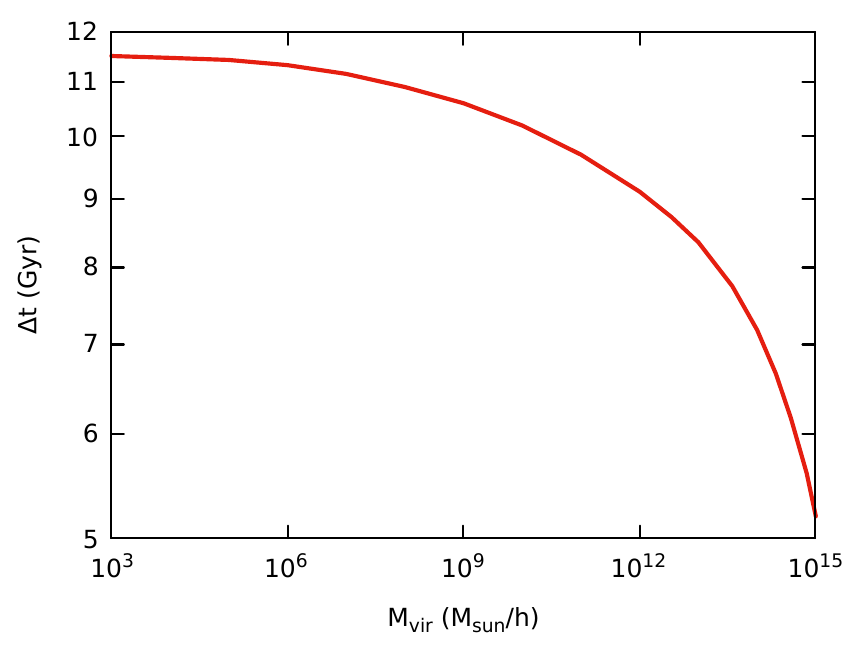}
		\caption{Time after the halo virialization, $\Delta t$, as a function of halo virial mass, $M_{\rm vir}$. The unit of time is Gyr, which can be translated to $t_{0}$.}
		\label{fig:t_age}
	\end{minipage}
\end{figure}

Furthermore, the calculation of the time after the halo virilization as a function of halo mass is the other important factor that plays a crucial role in the evolution of the concentration parameter. In other words, the time after the halo virialization varies from halo to halo. This means that one has to consider the average age of halos in terms of those masses in the calculation. To calculate the halo age as a function of the halo mass, one can define the critical overdensity of a halo with present-time virial mass $M_{\rm vir}$ as a function of the formation redshift $z_{\rm f}$~\cite{Ludlow:2013vxa}
\begin{equation}\label{delta_c_z_f}
\delta_{\rm c}(z_{\rm f})=\frac{\delta_{c}}{D(z_{\rm f})}=\delta_{\rm c}+0.447\sqrt{2[\sigma^{2}(f M,0)-\sigma^{2}(M,0)]},
\end{equation}
where $f=0.068$ is a fitting parameter specified from the accretion histories. Using the growth factor definition and applying Eqs.~(\ref{sigma_m_vir}) and (\ref{m_vir}) for $\sigma(M, z)$, one can solve Eq.~(\ref{delta_c_z_f}) for the formation redshift, which leads to the definition of the time after the halo virialization~\cite{Bondarenko:2017rfu},
\begin{equation}
\Delta t = t_{\rm univ} - t(z_{\rm f}),
\end{equation}
where $t_{\rm univ}$ is the current age of the Universe and
\begin{equation}
t(z_{\rm f})=\frac{1}{H_{0}}\int_{z_{\rm f}}^{\infty}\frac{dz}{(1+z)\sqrt{\Omega_{\rm m}(1+z)^{3}+\Omega_{\Lambda}}}.
\end{equation}
In Fig.~\ref{fig:t_age} the time after the halo virialization is plotted as a function of virial mass. As can be seen from the figure, the time elapsed from the halo virialization changes inversely with halo mass. In other words, smaller halos are older than larger halos because they have already become virialized. Therefore, subhalos have a higher concentration, which is well incorporated into Eq. (\ref{M_c_t}). It should be noted that the unit of $\Delta t$ is Gyr, which can easily be translated to $t_{0}$.

\subsection{Halo mass function}\label{subsec:hmf}
Gravitational collapse is a suitable and crucial frame for modeling spherically symmetric overdensities that lead to the formation of virialized dark matter halos. Therefore, having a suitable model based on spherically symmetric gravitational collapse is necessary to describe the behavior of dark matter halos and to classify them based on their mass distribution. In this regard, a function called the halo mass function has been introduced in cosmology that provides a convenient description of the mass distribution of dark matter halos \cite{Reed:2006rw, Lukic:2007fc, Murray:2013sna}. 

The halo mass function is a powerful probe in cosmology and related theories. Hence, having a proper halo mass function with high prediction accuracy is an essential tool, and can be used as the initial assumption of simulations related to the formation of cosmic structures. In other words, the halo mass function describes structures whose densities exceed the threshold, decouple from the cosmological expansion, and lead to gravitational collapse. 

In cosmology, one can introduce a parameter called the density contrast, i.e., $\delta(x)\equiv[\rho(x)-\bar{\rho}]/\bar{\rho}$, where $\rho(x)$ is the local density at an arbitrary point $x$ and $\bar{\rho}$ is the average background energy density. The density contrast is a criterion for the local increase of density fluctuations. For this reason, it can indicate the conditions under which the structures will form.

As mentioned, a threshold value for the density contrast has been calculated in cosmology for spherical-collapse halo models that is equal to $\delta_{\rm c} =1.686$. This value is independent of all local quantities such as mass and radius, and depends only on the redshift in a way that, in a narrow redshift range, it can be considered as a constant threshold \cite{Lukic:2007fc}. 

On the other hand, to characterize various fits for dark matter halos, a convenient definition of the differential halo mass function was introduced in Ref.~\cite{Jenkins:2000bv} as
\begin{equation} \label{mf}
 \frac{dn}{dM}=f(\sigma)\frac{\rho_{\rm m}}{M}\frac{d\ln(\sigma^{-1})}{dM},
\end{equation}
where $n(M)$ is the number density of dark matter halos, $M$ is the halo mass, $\rho_{\rm m}$ is the cosmological matter density and $f(\sigma)$ is a function that is related to the geometrical conditions for the overdensities at the collapse time which can be derived from the mathematical approaches or numerical simulations.

Many studies have been performed to obtain an appropriate halo mass function with the aim of accurate predictions that can provide the best fit for cosmic observations. One of the most successful models introduced for the halo mass function was created by Press and Schechter~\cite{ps}. Their formalism was based on an analytical approach assuming the homogeneous and isotropic gravitational collapse of overdensities. Under these assumptions, they presented a suitable halo mass function as
\begin{equation}\label{mf1}
 f_{\rm P\mbox{-}S}(\sigma) = \sqrt{\frac{2}{\pi}}\frac{\delta_{\rm sc}}{\sigma}\exp\left(\frac{-\delta_{\rm sc}^{2}}{2\sigma^{2}}\right),
\end{equation}
which is known as the Press-Schechter (P-S) halo mass function. Such a formalism predicts how many dark matter halos could exist between the masses $M$ and $M+dM$, assuming that the hierarchical structure formation occurs in objects with masses larger than $M$. In addition to the P-S formalism, other models have been proposed for the halo mass function based on numerical simulations and analytical approaches \cite{Jenkins:2000bv, st, Reed:2003sq, Warren:2005ey, Reed:2006rw}. 

In this work, we consider the shape of halos to be spherical. This choice is well justified by direct effects of the self-interaction among dark matter particles, as simulations indicate that SIDM halos are more spherical than CDM ones~\cite{Brinckmann:2017uve}. A broad research on the SIDM halo shapes and comparing those with the observational data has been conducted in Ref.~\cite{Peter:2012jh}. The result reveals that SIDM halo models with higher values of $\sigma/m$ are rounder than CDM halo models, particularly in their inner regions. The observed halo shape varies with respect to the various contributors, such as the halo size and the definition of the halo shape. In fact, one of the observational constraints on the value of  $\sigma/m$ could be found through the shape of halos. For $\sigma/m > 1~\rm cm^{2}g^{-1}$, the spherical assumption is a natural consequence of SIDM halo models. Under these considerations, we use the P-S mass function to calculate the merger rate of PBHs in SIDM halo models. 

Up to now, we have specified the framework for SIDM halo models. Hence, we are able to study the merger rate of PBHs considering SIDM halo models. For this purpose, in the following section we will discuss the encounter condition of PBHs in the medium of dark matter halos, their binary formation conditions, and their merger rates.
\section{Merger Rate of PBHs}
\label{sec:iii}
\subsection{Merger rate of PBHs within each halo} \label{sec.iiia}
In this section we calculate the merger rate of PBHs in the framework of SIDM halo models. PBHs are a special type of black hole that follow a different process of formation compared to the formation of black holes of astrophysical origin. Sufficiently dense regions in the early Universe may lead to the formation of PBHs shortly after the big bang due to the direct collapse of overdensities that exceed their thresholds. In addition, the random distribution of PBHs in dark matter halos allows them to form binaries not only during the radiation-dominated era but also in the late-time Universe.

As mentioned, it is believed that the detection of gravitational waves from a binary black hole merger via the LIGO-Virgo detectors would be compatible with the merger rate of PBHs with a typical mass $30~M_{\odot}$, if a significant fraction of dark matter is formed by PBHs. In this work, we propose to interpret the conditions under which stellar-mass  PBHs (as a proposed candidate for dark matter) in the medium of dark matter halos could encounter each other, form binaries, and eventually merge. In fact, our aim is to calculate the merger rate of PBHs in SIDM halo models, and to compare it with the corresponding result of CDM halo models.

Consider two PBHs with masses $m_{1}$ and $m_{2}$ and relative velocity at large separation $v_{\rm rel}=|v_{1}-v_{2}|$ that suddenly encounter each other in a dark matter halo. Due to the maximum scattering amplitude, significant gravitational radiation would happen when they are located at the closest separation (i.e., at periastron) from each other. The time-average gravitational energy emitted from such an encounter can be calculated in the context of Keplerian mechanics \cite{peters} as
\begin{equation} \label{energy1}
	\langle\frac{dE_{\rm rad}}{dt}\rangle = -\frac{32}{5}\frac{G^{4}(m_{1}m_{2})^{2}(m_{1}+m_{2})}{c^{5}a^{3/2}r_{\rm p}^{7/2}(1+e)^{7/2}}\left(1+\frac{73}{24}e+\frac{37}{96}e^{2} \right),
\end{equation}
where $G$ is the gravitational constant, $c$ is the velocity of light, $a$ and $e$ are the semimajor axis and eccentricity of the orbit, and $r_{\rm p}=a(1-e)$ is the periastron. As can be inferred from Eq.~(\ref{energy1}), the evolution of formed binaries depends on the properties of these orbital parameters. Near the periastron, the trajectory can be roughly considered as an unperturbed parabolic orbit that corresponds to an ellipse with the highest eccentricity (i.e., $e = 1$), as the strong gravitational limits dominate this binary in such a way that the most gravitational radiation occurs at this point. Under these assumptions, one can calculate the radiated gravitational energy after one orbital period as
\begin{equation}\label{energy2}
	\Delta E_{\rm rad}=\frac{85 \pi }{12\sqrt{2}}\frac{(m_{1}m_{2})^{2}\sqrt{(m_{1}+m_{2})}}{c^{5}r_{p}^{7/2}}.
\end{equation}
If the radiated gravitational energy is greater than the kinetic energy of the PBHs, then they will become gravitationally bound and form a binary. This condition leads to a maximum value for the periastron as
\begin{equation}\label{periast}
	r_{\rm p, max} = \left[\frac{85 \pi}{6\sqrt{2}}\frac{G^{7/2}m_{1}m_{2}(m_{1}+m_{2})^{3/2}}{c^{5}v_{\rm rel}^{2}}\right]^{2/7}.
\end{equation}
This maximum value implies that the PBHs would be able to form a binary if the condition $r_{\rm p}<r_{\rm p, max}$ is satisfied. On the other hand, in the Newtonian approximation, the relation between the impact parameter $b$ and the periastron can be obtained as \cite{Sasaki:2018dmp}
\begin{equation}\label{impact}
	b(r_{\rm p}) = \frac{2G(m_{1}+m_{2})r_{\rm p}}{v_{\rm rel}^{2}} + r_{\rm p}^{2}.
\end{equation}
Moreover, for such an encounter, the cross section for the binary formation $\xi(m_{1}, m_{2}, v_{\rm rel})$ is equal to the area of a circle with a radius of $b(r_{\rm p, max})$~\cite{quinlan,Mouri:2002mc}.

In this work, we are interested in studying the merger event rate of PBHs that are consistent with the mergers obtained via the LIGO-Virgo detectors, i.e., $30~M_{\odot}\mbox{-}30~M_{\odot}$ events in dark matter halos. For this purpose, we consider binaries with equal-mass components, i.e., $m_{1}=m_{2}=M_{\rm PBH}$, and set $v_{\rm rel}=v_{\rm PBH}$. Under these assumptions and considering the strong limit of gravitational focusing (i.e., $r_{\rm p}\ll b$), the cross section for binary formation $\xi$ can be obtained as
\begin{multline}\label{cross}
	\xi \simeq 4\pi \left(\frac{85\pi}{3}\right)^{2/7}\left(\frac{M_{\rm PBH}^{2}G^{2}}{c^{10/7}v_{\rm PBH}^{18/7}}\right) \\
\kern 0.5pc	\simeq 1.37 \times 10^{-14}\left(\frac{M_{\rm PBH}}{30M_{\odot}}\right)^{2}\left(\frac{v_{\rm PBH}}{200\,\rm km/s}\right)^{-18/7} \rm in\hspace{0.2cm}(pc)^{2},
\end{multline}
where, in the last line, we have normalized the PBH mass to $30~M_{\odot}$ and the PBH relative velocity to average velocities of dark matter halos, i.e., $200~\rm km/s$. Accordingly, the rate of PBH binary formation within each halo can be specified using the relation \cite{bird, Nishikawa:2017chy}
\begin{equation}\label{eq:per}
	\Gamma = 2\pi\int_{0}^{r_{\rm vir}} r^{2}\left(\frac{\rho(r)\,f_{\rm PBH}}{M_{\rm PBH}}\right)^{2}\langle \xi v_{\rm PBH}\rangle dr,
\end{equation}
where $0<f_{\rm PBH}\leq1$ is the fraction of PBHs, which indicates those contribution to dark matter, $\rho(r)$ is the halo density profile, and the angle bracket shows an average over the PBH relative velocity distribution in the galactic halo. As mentioned in Sec. \ref{sec.i}, even if a small fraction of dark matter is strongly self-interacting, and the rest of it is the other component, the model can remove astrophysical problems. Consequently, unlike the CDM halo model, considering that all dark matter is composed of PBHs is not consistent with the essence of the SIDM model. As the contribution of self-interacting dark matter particles, even if it is small, must always be maintained within the SIDM model, we have left out $f_{\rm PBH}=1$ from the results of the SIDM model.

\begin{figure}[t!]
	\begin{minipage}{1\linewidth}
		\includegraphics[width=1\textwidth]{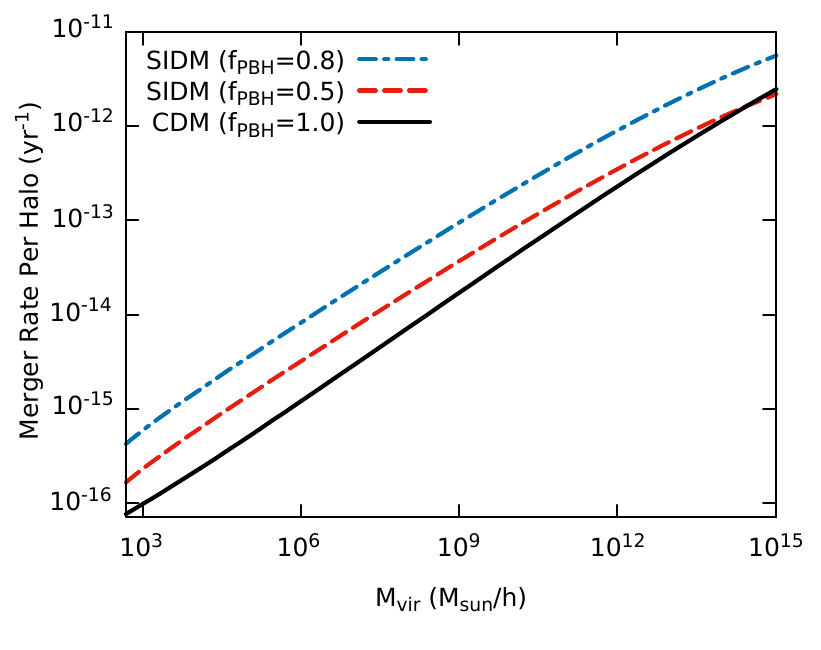}
		\caption{PBH merger rate within each halo for the SIDM and CDM halo models in the present-time Universe. The dot-dashed (blue) and dashed (red) lines represent the merger rate of PBHs in SIDM halo models considering $f_{\rm PBH}=0.8$ and $0.5$, respectively, whereas the solid (black) line shows CDM halo models with the NFW density profile for $f_{\rm PBH}=1.0$.}
		\label{fig:per}
	\end{minipage}
\end{figure}

In addition, the mass bounded via the virial radius of the halo, the virialized mass, can be calculated as
\begin{equation}\label{2a}
	M_{\rm vir}=\int_{0}^{r_{\rm vir}}4\pi r^{2}\rho(r) dr.
\end{equation}
Another important factor in calculating the merger rate of PBHs is the halo velocity dispersion. It should be noted that the self-interacting scenario of dark matter only changes the velocity dispersion in the inner region of a halo and has no significant effect in outward radii, where the velocity dispersion reaches the maximum value \cite{Robertson:2018anx}. Therefore, with a plausible approximation, one can use the relation obtained in Ref.~\cite{prada} for the halo velocity dispersion, namely,
\begin{equation}
	v_{\rm disp}=\frac{v_{\rm max}}{\sqrt{2}}=\sqrt{\frac{GM(r<r_{\rm max})}{r_{\rm max}}}.
\end{equation}
We also demand that the relative velocity distribution of PBHs in the galactic halo corresponds to the Maxwell-Boltzmann statistics with the following probability distribution function while considering a cutoff at the halo virial velocity:
\begin{equation}\label{3}
	P(v_{\rm PBH}, v_{\rm disp})=F_{0}\left[\exp\left(-\frac{v_{\rm PBH}^{2}}{v_{\rm disp}^{2}}
	\right)-\exp\left(-\frac{v_{\rm vir}^{2}}{v_{\rm disp}^{2}}\right)\right],
\end{equation}
where $F_{0}$ is specified by $4\pi \int_{0}^{v_{\rm vir}} P(v)v^{2}dv=1$. It is clear from Eqs.~(\ref{early_time}), (\ref{late_time}), and (\ref{eq:per}) that the concentration parameter plays a crucial role in calculating the merger event rate of PBHs within each halo. Hence, due to the fact that this parameter for SIDM halo models deviates significantly from that for CDM halo models, it can be expected that the merger event rate of PBHs within each halo for SIDM halo models varies from that obtained from CDM halo models. Accordingly, in order to calculate the merger event rate of PBHs for SIDM models, we use Eq.~(\ref{M_c_t}) for the concentration parameter and employ Eq.~(\ref{m_c_NFW}) to obtain the corresponding result for CDM models. We also set the mass of PBHs to be $30\,M_{\odot}$. Moreover, as mentioned in Sec.~\ref{sec.iia}, one can expect that the evolution of the density profile remains the same for all dark matter halos in the SIDM model with various masses after $\Delta t=100\,t_0$, or even at other times during the second epoch. Hence, to calculate the merger event rate of PBHs per halo we employ Eq.~(\ref{late_time}) as the density profile in SIDM halo models.

In Fig.~\ref{fig:per}, we show the merger event rate of PBHs within each halo with respect to the halo mass and the PBH fraction (i.e., $f_{\rm PBH}<1$) for SIDM models, and compare them with the corresponding result for CDM models with $f_{\rm PBH}=1$. It should be considered that these calculations are performed for the present-time Universe. As mentioned earlier, considering that all dark matter is made of PBHs is not compatible with the SIDM model. On the other hand, it can be easily realized from Eq.~(\ref{eq:per}) that the PBH merger rate within each halo changes directly with the square of $f_{\rm PBH}$. The results show that the merger rate of PBHs per halo for SIDM models with $f_{\rm PBH}>0.32$ is higher than that obtained from CDM models with $f_{\rm PBH}=1$.

The main reason for the difference in the merger rate between the two models is the time evolution of the halo density profile in the SIDM halo models. As can be seen in Fig.~\ref{figure_1}, the density profile inside the inner region of SIDM halos becomes cored until $\Delta t = 25 \, t_0$ after the virialization time, and it is lower than that in the CDM halos during the same time. On the contrary, the density in the central region increases during $\Delta t > 25 \, t_0$ and at later epoch the core collapses. At this stage, the inner region of the SIDM halos is much denser than the CDM halos. In addition, it is clear from Eq.~(\ref{eq:per}) that the merger event rate of PBHs is directly proportional to the density profile. Thus, as expected, the time evolution of the density profile resulting from SIDM halo models in the inner regions of halos leads to a modification of the merger rate of PBHs residing in dark matter halos.

In this study, we have focused on PBH binaries formed in dark matter halos in the present-time Universe. However, it is possible that a large number of PBHs, due to the initial clustering and high probability of proximity, could also have decoupled from the Hubble flow, gravitationally bound and formed binaries in the early Universe \cite{Ioka:1998nz}. PBH binaries formed in the early Universe continuously emit gravitational waves, gradually shrink, and eventually merge. However, some of them would disrupt, due to the tidal forces of surrounding PBHs, before those mergers \cite{Ali-Haimoud:2017rtz, Raidal:2018bbj, Kavanagh:2018ggo}. On the other hand, the orbital parameters of binaries play an essential role in the time to merge. Hence, due to the random distribution of orbital parameters of binaries in the early Universe, some binaries have already merged, some others merge in the present-time Universe, and others will merge in the future. As a result, those PBH binaries that are supposed to merge in the present-time Universe will dramatically increase the merger rate today. It has been claimed that in this mechanism of PBH binary formation, to justify the LIGO-Virgo observations, PBHs must constitute a very small fraction of dark matter \cite{Sasaki:2016jop}.

In the following, we intend to study the effect of SIDM halo models on the total merger rate of PBHs per unit time and per unit volume and compare it with the corresponding results of CDM halo models.
\subsection{Total merger rate of PBHs }
\subsubsection{Present-time Universe}
It should be noted that the accumulation of binary black hole merger rates can be deduced from the data recorded by gravitational-wave detectors. The quantities we have explained so far are needed to obtain the cumulative merger rate of PBHs per unit time and per unit volume. Hence, to calculate the total merger rate of PBHs, the last step is to convolve the halo mass function $dn/dM_{\rm vir}$ with the merger rate per halo $\Gamma(M_{\rm vir})$. Under these considerations, the total merger rate can be obtained as
\begin{equation}\label{tot_mer}
	\mathcal{R}=\int_{M_{\rm min}}\frac{dn}{dM_{\rm vir}}\Gamma(M_{\rm vir})dM_{\rm vir}.
\end{equation}
We demand that the initial conditions for the formation of both models, i.e., SIDM and CDM halo models, are the same based on a spherically symmetric gravitational collapse. Therefore, to obtain the total merger rate of PBHs, we use the P-S halo mass function introduced in Sec.~\ref{subsec:hmf}. Due to the presence of the exponential term in the P-S halo mass function, the upper limit of the halo mass has no significant effect on the final result, whereas the role of the lower limit of the halo mass is crucial. Because, it has been indicated that the subhalos must contain dark matter with lower velocity dispersion and higher density than the larger host halos \cite{bird, Moore:1999nt, Kamionkowski:2008vw}. In this regard, it is expected from Eq.~(\ref{eq:per}) that the smallest halos provide a significant contribution to the merger rate of PBHs.

\begin{figure}[t!]
	\begin{minipage}{1\linewidth}
		\includegraphics[width=1\textwidth]{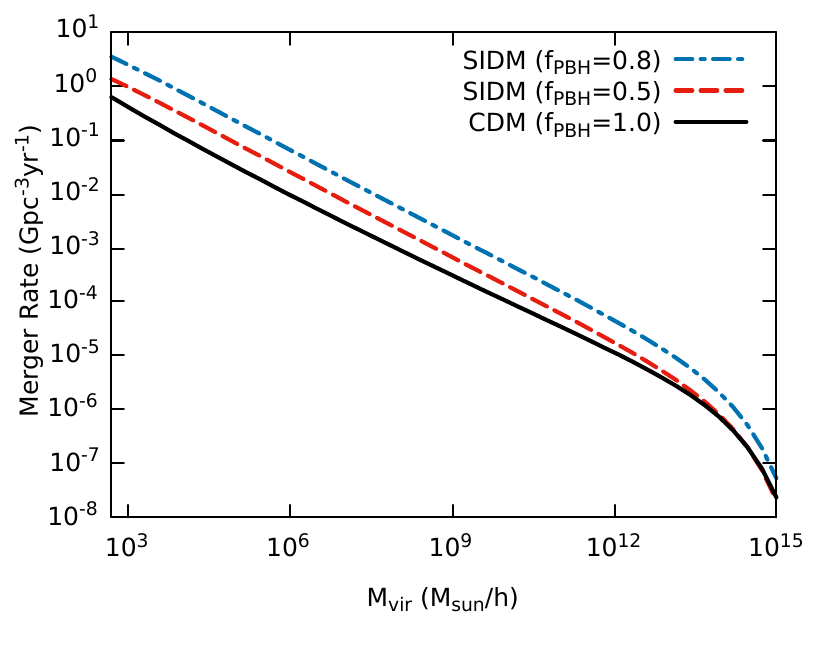}
		\caption{Total merger event rate of PBHs per unit source time and per unit comoving volume for SIDM and CDM halo models over the halo virial mass in the present-time Universe. The dot-dashed (blue) and dashed (red) lines represent the merger rate of PBHs in SIDM halo models while considering $f_{\rm PBH}=0.8$ and $0.5$, respectively, whereas the solid (black) line shows CDM halo models with the NFW density profile for $f_{\rm PBH}=1.0$.}
		\label{fig:tot}
	\end{minipage}
\end{figure}

It is believed that the smallest halos evaporate faster than larger halos because they have already become virialized. On the other hand, the time scale of halo evaporation depends on the number of independent objects that may reside in the halos (i.e., $N=M_{\rm vir}/M_{\rm PBH}$). It has been shown~\cite{bird} that the evaporation time of halos with a mass of $400\,M_{\odot}$, which includes PBHs with a mass of $30\,M_{\odot}$, is about $3~\rm Gyr$. On the other hand, halo evaporation during the matter-dominated era is compensated by some processes, such as the accretion of surrounding materials into the halo or the merging of smaller halos. However, the compensating processes slow down during the dark-energy-dominated era (i.e., approximately $3~\rm Gyr$ ago) due to the accelerating expansion of the Universe. As a result, it can be assumed that signals from the halos with an evaporation time of less than $3~\rm Gyr$ are negligible. Hence, one can ignore the signal from halos with masses less than $400\,M_{\odot}$ (see, e.g., Refs.~\cite{bird, Fakhry:2020plg} for more details). With this argument, we set the lower limit of the halo mass to be $400\,M_{\odot}$, while containing PBHs with a typical mass of $30\,M_{\odot}$.

In Fig.~\ref{fig:tot} we indicate the merger rate of PBHs per unit time and per unit volume for SIDM halo models as a function of virial mass and PBH fraction (i.e., $f_{\rm PBH}<1$), and compare it with the corresponding one obtained from CDM halo models with $f_{\rm PBH}=1$. Note that these calculations are also performed for the present-time Universe. As can be understood from Eqs.~(\ref{eq:per}) and (\ref{tot_mer}), the total merger rate of PBHs changes directly with the square of $f_{\rm PBH}$. In this regard, the results indicate that the merger rate of PBHs for SIDM halo models with $f_{\rm PBH}>0.32$ is higher than that extracted from CDM halo models even with $f_{\rm PBH}=1$. In other words, even considering the $100\%$ contribution of PBHs in dark matter in the CDM model, one would potentially expect to have an enhancement in the total merger rate of PBHs in the SIDM model, if at least $32\%$ of dark matter is made of PBHs. Moreover, Fig.~\ref{fig:tot} shows that the merger rate of PBHs for both models decreases with increasing halo mass. This result is due to the presence of the exponential term in the halo mass function that well justifies the inverse evolution of the dark matter density and the direct evolution of the dark matter velocity dispersion with the halo mass. The total merger rate has been calculated by integrating over the surface below the curves.

\subsubsection{Redshift evolution of PBH merger rate} \label{sec:iiib}
The history of the black hole merger rate during the evolution of the Universe is one of the suitable criteria to separate black hole formation scenarios \cite{Sasaki:2018dmp}. Moreover, the development of instruments and improvement of their accuracy can enable gravitational-wave detectors to probe events at higher redshifts. Nowadays, the Advanced LIGO (aLIGO)-Advanced Virgo (aVirgo) detectors can detect binary mergers up to $z\sim0.75$, which approximately corresponds to a comoving volume around $50~\rm Gpc^{3}$ \cite{LIGOScientific:2018jsj, Abbott:2020niy}. For this reason, we are going to discuss the evolution of the PBH merger rate as a function of redshift for SIDM and CDM halo models. In this regard, it is obvious that Eq.~(\ref{tot_mer}) depends on the redshift through the halo mass function and the concentration parameter~\cite{Mandic:2016lcn}.

\begin{figure}[t!]
	\begin{minipage}{1\linewidth}
		\includegraphics[width=0.97\textwidth]{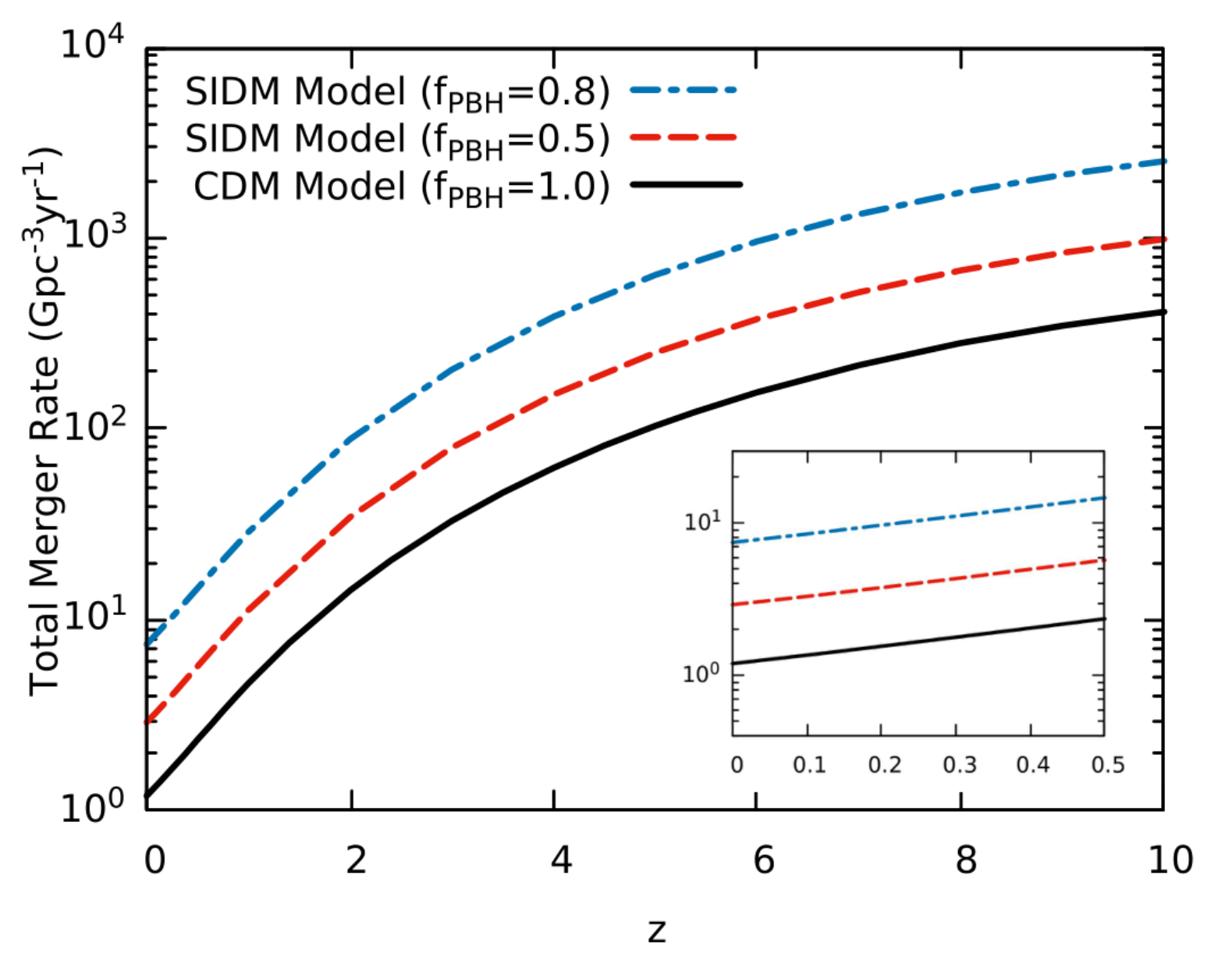}
		\caption{Redshift evolution of the total merger rate of PBHs per unit source time and per unit comoving volume for SIDM and CDM halo models. The dot-dashed (blue) and dashed (red) lines represent the merger rate of PBHs in SIDM halo models while considering $f_{\rm PBH}=0.8$ and $0.5$, respectively, whereas the solid (black) line shows the corresponding result for CDM halo models with the NFW density profile for $f_{\rm PBH}=1.0$.}
		\label{fig:red}
	\end{minipage}
\end{figure}

Fig.~\ref{fig:red} shows the redshift evolution of the total merger rate of PBHs for the SIDM halo model with two values of the PBH fractions (i.e., $f_{\rm PBH}=0.5$ and $0.8$) and compares them with that obtained from the CDM halo model with $f_{\rm PBH}=1$. Although the main purpose of this work is to calculate the merger rate of PBHs in the late-time Universe in a way that it can be compared with the mergers recorded by the LIGO-Virgo detectors, to provide theoretical predictions of the models proposed in this work for the future of gravitational waves and the development of those detectors, we show the evolution of the merger rate of PBHs including the redshift information up to $z\simeq 10$. As can be seen from the figure, both models predict that the total merger rate of PBHs is directly proportional to redshift. In other words, this means that the PBHs are more likely to form binaries at higher redshifts than in the present-time Universe, which is compatible with other studies~\cite{Fakhry:2020plg, Sasaki:2018dmp, Mandic:2016lcn, Gow:2019pok}. In addition, for mergers related to the late-time Universe and comparable with the current sensitivity of the LIGO-Virgo detectors, we display the merger rate of PBHs up to redshift $z=0.5$ as an inset figure. Also, the results demonstrate that the total merger rate of PBHs for SIDM halo models (with $f_{\rm PBH}>0.32$) is higher than that calculated for CDM halo models. As a result, it can be inferred that the merger rate of PBHs will be amplified over time if SIDM halo models are reliable. In other words, due to the time evolution of the density profile, the merger rate of PBHs in SIDM models evolves in a way that is quite different than the evolutionary behavior of the merger rate of PBHs derived from CDM models. This result can potentially be validated as a distinguishing feature between SIDM and CDM halo models. Moreover, the results obtained in this study for the CDM halo model, especially at lower redshifts, are in good agreement with the previous calculations \cite{bird, Fakhry:2020plg, Mandic:2016lcn}. However, uncertainties in the local density distributions in CDM halo models \cite{Green:2017odb}, and uncertainties in the shape of the halo mass functions and collapse conditions \cite{Wu:2009we}, which increase at higher redshifts \cite{Mandic:2016lcn}, may affect the merger rate of PBHs.

As mentioned in Sec. \ref{sec.iiia}, in these calculations we assume that the gravitational focusing term dominates. Under this assumption, dissipation by gravitational radiation is much more important than that by tidal forces \cite{quinlan}. In other words, once the binary is formed, one can require that the time until the PBH binary merges is less than a Hubble time \cite{bird}. Also, it is known that the merger time of a binary black hole is a function of velocity dispersion of halos (from hours to kiloyears) \cite{OLeary:2008myb}. Thus, compared to cosmological time scales, such mergers seem instantaneous. Hence, due to the instantaneous time to merge, the disruption of PBH binaries via surrounding PBHs is expected to be rare. In addition, we assume that the merger rate of binaries consisting of nondissipative three-body encounters must be negligible, because these often lead to wide binaries that do not have enough binding energy to merge during the age of the Universe~\cite{quinlan}. As a result, their mergers cannot be recorded by the LIGO-Virgo detectors. Although the method presented in this work yields a suitable framework for the calculation of the PBH merger rate in galactic halos, its assumptions are not necessarily satisfied and some perturbing processes such as the disruption of PBH binaries caused by the surrounding PBHs remain as uncertainties \cite{Ali-Haimoud:2017rtz, Raidal:2018bbj, Vaskonen:2019jpv}. Specifically, these uncertainties may be large in the SIDM model, because the enhancement of the density in the SIDM model can increase the probability of a close encounter with surrounding PBHs, which may disrupt the binary before it merges.

\subsubsection{Constraint on PBH fraction}
The study of constraints arising from the effects of PBHs on the observable Universe has always been one of the main topics discussed in the literature. The importance of PBH constraints is that they can provide a clear picture of the number density of PBHs and their contribution to dark matter. In addition to all of the observational constraints that have been placed on the abundance of PBHs, the merger rate of these black holes can potentially lead to a constraint on the PBH fraction in dark matter. 

Nowadays, most of the PBH mass ranges have been constrained due to the various cosmological processes \cite{Carr:2020gox}. There is only a small window, known as the asteroid-mass PBHs \cite{Katz:2018zrn, Montero-Camacho:2019jte, Smyth:2019whb, coogan, Ray:2021mxu, Picker:2021jxl}, that is still open. This mass range of PBHs could potentially contain a significant fraction of dark matter. Fortunately, the black hole merger events recorded by the LIGO-Virgo detectors are a convenient and accessible criterion to evaluate the validity of PBH mergers generated by various halo models. On the other hand, comparing the PBH merger rate obtained from halo models with the mergers reported by the LIGO-Virgo detectors leads to a constraint on the fraction of PBHs. Note that the constraints arising from the merger rate of PBHs in theoretical models are the upper limits that are allowed by the gravitational-wave detectors since the black hole mergers with astrophysical origins are also likely to be recorded by the LIGO-Virgo detectors.

\begin{table*}[t] 
	\caption{Total merger event rate of PBHs with different fractions and masses for SIDM and CDM halo models. The results are related to the present-time Universe.}
	\centering 
	\begin{tabular}{| c | c | c | c |}
		\hline 
		\hline 
		Halo Model & $f_{\rm PBH}$ & $M_{\rm PBH}$ ($M_{\odot}$) & Total merger rate~$\rm (Gpc^{-3}yr^{-1})$\\ [0.5ex]
		\hline
			&	 		& $10$ &$0.20$\\	
			& $0.1$ & $30$ &$0.11$\\	
			&	 		& $100$ &$0.07$\\	 \cline{2-4}
			
			&		   & $10$ &$5.19$\\	
	SIDM	&$0.5$ & $30$ &$2.92$\\	
			&		   & $100$ &$1.83$\\	\cline{2-4}
			&		   & $10$ &$13.31$\\	
			&$0.8$ & $30$ &$7.48$\\	
			&		   & $100$ &$4.96$\\	
		\hline
			&		   & $10$ &$2.43$\\	
	CDM	&$1.0$ & $30$ &$1.21$\\	
			&		   & $100$ &$0.71$\\	
		\hline 
		\hline
	\end{tabular}
	\label{table:info2}
\end{table*}

On the other hand, according to the criterion related to the evaporation time of dark matter halos containing PBHs, one can constrain the smallest halos that have not yet evaporated by the present-time Universe. In fact, the lower limit of the halo mass directly depends on the mass of PBHs located in the host halos (see, e.g., Ref. \cite{Fakhry:2020plg}). The main idea is to find out whether the merger rate and the fraction of PBHs modify with their smaller or larger masses than $30\,M_{\odot}$. Under these assumptions, the merger rate for the various masses of PBHs can be calculated. In Table~\ref{table:info2}, we show the total merger rate of PBHs for several masses of PBHs while considering SIDM and CDM halo models. We also consider some values of PBH fractions when considering SIDM halo models for comparison. As is clear, the merger rate of PBHs changes inversely with their masses, and is directly proportional to $f_{\rm PBH}^{2}$.

\begin{figure}[t!]
	\begin{minipage}{1\linewidth}
		\includegraphics[width=1\textwidth]{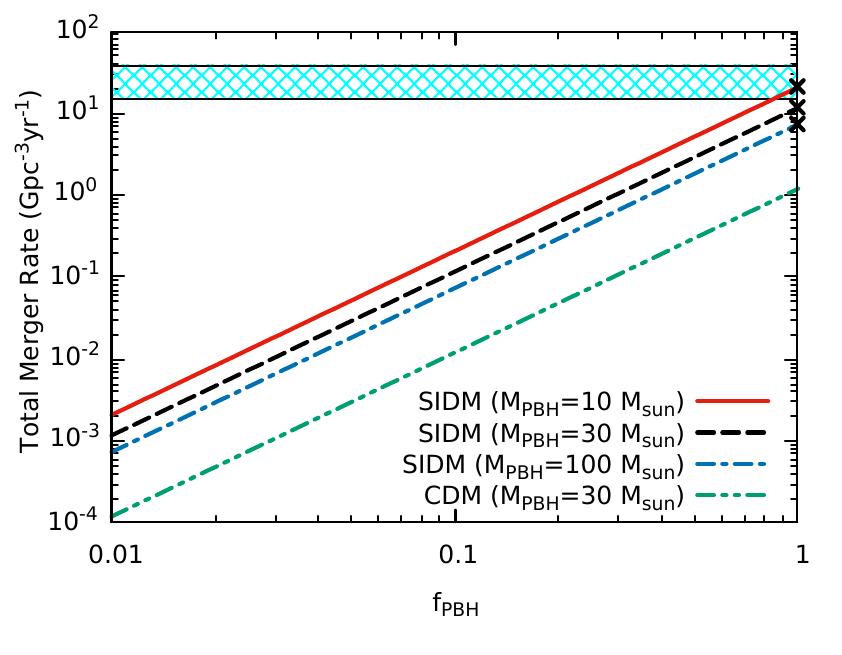}
		\caption{Total merger event rate of PBHs for SIDM and CDM halo models with respect to the PBH fraction and mass. The solid (red), dashed (black), and dot-dashed (blue) lines show this relation for SIDM halo models considering a PBH mass of $M_{\rm PBH} = 10$, $30$, and $100\,M_{\odot}$, respectively, while the dot-dot-dashed (green) line indicates that corresponding to CDM halo models with $M_{\rm PBH} = 30\,M_{\odot}$. The shaded (cyan) band is the total merger rate of black holes estimated by the aLIGO-aVirgo detectors during the third observing run, i.e., $15.3\mbox{-}38.8\,\rm Gpc^{-3} \rm yr^{-1}$. Note that the results for SIDM halo models are not strictly valid for $f_{\rm PBH} \simeq 1$.} 
		\label{fig:ab}
	\end{minipage}
\end{figure}

\begin{figure}[t!]
	\begin{minipage}{1\linewidth}
		\includegraphics[width=1\textwidth]{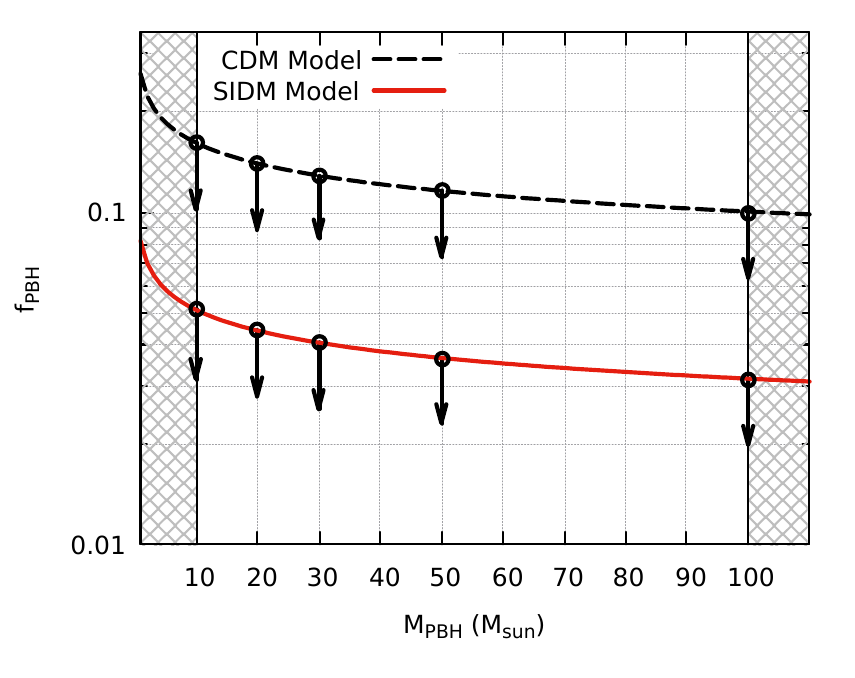}
	\\ \hspace*{0.5cm} \\
		\includegraphics[width=1\textwidth]{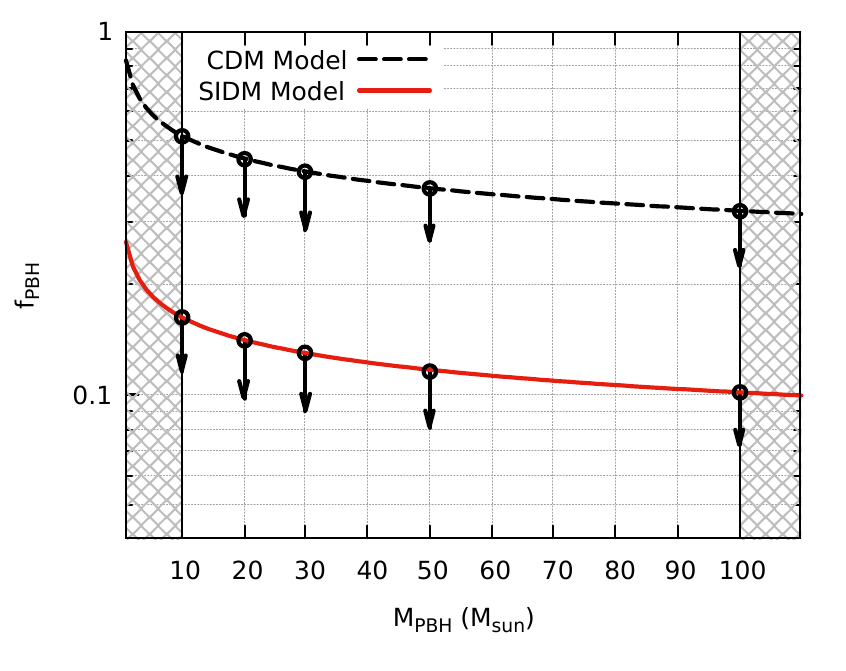}
		\caption{Expected upper bounds on the fraction of PBHs, $f_{\rm PBH}$, as a function of their masses, $10\,M_{\odot} < M_{\rm PBH} < 100\,M_{\odot}$ and considering SIDM and CDM halo models. Top: calibrated for the situation in which one should expect to detect at least one $30\,M_{\odot}\mbox{-}30\,M_{\odot}$ event in the comoving volume $50\,\rm Gpc^{3}$ annually. Bottom: calibrated for the situation in which one should expect to detect at least ten $30\,M_{\odot}\mbox{-}30\,M_{\odot}$ events in the same comoving volume. The solid (red) lines indicate the results for SIDM halo models, while the dashed (black) lines show that extracted from CDM halo models.}
		\label{fig:f-m}
	\end{minipage}
\end{figure}

In Fig.~\ref{fig:ab} we show the total merger rate of PBHs with respect to their fraction and masses while considering the SIDM halo models. For comparison, we also show the corresponding results for CDM halo models with $M_{\rm PBH}=30\,M_{\odot}$. As can be seen from the figure, we have left out $f_{\rm PBH}=1$ from our calculations with the SIDM halo models. Because the results are not strictly valid for the situation under which the total dark matter is made of PBHs. In this figure, the shaded band indicates the total merger rate of black holes recorded by the aLIGO-aVirgo detectors during the third observing run, i.e., $15.3\mbox{-}38.8\,\rm Gpc^{-3}yr^{-1}$ \cite{Abbott:2020niy}.  The merger rate of PBHs changes inversely with their masses. Interestingly, despite all theoretical uncertainties, the merger rate of PBHs with a mass of $10\,M_{\odot}$ falls within the aLIGO-aVirgo merger range, whereas the corresponding result for the PBH mass ranges $10\,M_{\odot}<M_{\rm PBH}<100\,M_{\odot}$ will not be located in this window if CDM halo models are trusted. Furthermore, it should be noted that the shaded band specified by the aLIGO-aVirgo detectors includes stellar-mass black hole mergers with different masses. As an example, the aLIGO-aVirgo detectors recorded four merger events of black holes with a mass of about $10~M_{\odot}$ during the third observing run~\cite{Abbott:2020niy}, for which SIDM models predict an upper bound on the fraction of PBHs of $f_{\rm PBH}\sim\mathcal{O}(10^{-1})$. This means that the potential of dark matter can be dominated by a combination of SIDM particles and PBHs. These results show that the SIDM model can potentially improve the merger rate of PBHs and lead to stronger constraints compared to the results for the CDM model. However, uncertainties in the merger rate (e.g., those arising from the shape of the mass function) make it difficult to describe the population of PBHs at gravitational-wave observatories in a model-independent framework \cite{Lehmann:2020bby}.

Additionally, as mentioned, the aLIGO-aVirgo detectors can search for black hole merger events up to a comoving volume around $50\,{\rm Gpc^{3}}$. Roughly speaking, by considering the SIDM scenario indicated in Fig.~\ref{fig:ab}, the aLIGO-aVirgo detectors are expected to detect at least ten events annually with $f_{\rm PBH} > 10^{-1}$, whereas for the detection of at least one event during the same time, the PBH fraction is predicted to be $f_{\rm PBH} > 10^{-2}$.

As a final point, let us estimate how the constraints change with different PBH masses. For this purpose, according to the results obtained in Refs.~\cite{bird, Ali-Haimoud:2017rtz}, a relation between the fraction of PBHs and their masses can be estimated to be $f_{\rm PBH}\sim\left(M_{\rm PBH}/ 30\,M_{\odot}\right)^{-0.207}$, which is well consistent with the results obtained in this work.

In Fig.~\ref{fig:f-m} we show the expected upper bounds on the fraction of PBHs in terms of their masses, while considering SIDM halo models, and compare them with the corresponding calculations for CDM halo models. As can be seen, the fraction of PBHs is inversely proportional to their masses. The top plot represents the situation in which one can expect to record at least one $30\,M_{\odot}\mbox{-}30\,M_{\odot}$ event in the comoving volume $50\,{\rm Gpc^{3}}$ annually, while the bottom plot shows the corresponding results for at least ten $30\,M_{\odot}\mbox{-}30\,M_{\odot}$ events in the same comoving volume. Also, to represent the relative differences between the various results, several upper bounds on the fraction of PBHs obtained for some PBH masses are shown. The results indicate that SIDM models can potentially impose stronger constraints on the fraction of PBHs compared to the results of CDM models. Specifically, within the context of the SIDM halo models the fraction of PBHs should be of the order of $\mathcal{O}(10^{-2})$ if one expects to detect at least one $30\,M_{\odot}\mbox{-}30\,M_{\odot}$ event per year in the comoving volume $50~\rm Gpc^{3}$, while if at least ten $30\,M_{\odot}\mbox{-}30\,M_{\odot}$ events per year occur in the same comoving volume, the fraction of PBHs is predicted to be of the order of $\mathcal{O}(10^{-1})$.
\section{Conclusions} 
\label{sec:iv}
There are several candidates for dark matter that have been proposed in cosmology and particle physics. As dark matter makes up nearly $5$ times the contribution of baryonic matter, it would not be extraordinary to assume that dark matter itself includes a combination of various candidates. PBHs and SIDM are two such candidates. The idea of considering a proportion of dark matter with a non-negligible cross section per unit mass of particles and being able to interact with other dark matter particles can resolve many astrophysical problems, such as the missing satellite and core-cusp problems. On the other hand, the cutting edge of gravitational-wave detection in the LIGO-Virgo detectors has started a new era of cosmology in recent years. In particular, studying binary black hole mergers has rapidly been developed as nowadays we have access to direct observational data.

In this work, we have studied the merger rate of PBHs in SIDM halo models with a constant cross section per unit mass of dark matter particles. Although some constraints on these models indicate that a velocity-independent model of SIDM is unable to remove astrophysical problems, it is still worth studying SIDM models with a constant $\sigma/m$, as they could overlap with the velocity-dependent models with regard to particular assumptions. For this purpose, we have used the result of a previously performed simulation of SIDM with $\sigma/m=10~\rm cm^2 g^{-1}$ to specify a numerical description of the density profile of the SIDM halos as a function of time $\Delta t$ after the halo virialization. Then, to justify the evolution of the density profile in two different epochs (i.e., $\Delta t \leq t=25 \, t_0$ and $\Delta t > t=25 \, t_0$) after the halo virialization, we have proposed two relations for the density profile within the context of the SIDM halo models. In this regard, we have also found a relation for the concentration parameter, time, and virialized mass of a halo, which can justify the behavior of the evolution of the halo density related to SIDM halo models.

The density profile plays the central role in taking the effect of SIDM into account in the present analysis on the merger rate of PBHs. If the SIDM model is velocity dependent, the general effect of self-interactions on the density profile does not change. Ref.~\cite{Robertson:2018anx} describes the result of a simulation of three models of SIDM, including a velocity-dependent model (vdSIDM) and two models with fixed $\sigma/m$, which are SIDM0.1 with $\sigma/m=0.1~\rm cm^2 g^{-1}$ and SIDM1 with $\sigma/m=1~\rm cm^2 g^{-1}$ (see Fig. 2 therein). As can be seen, the effect of the vdSIDM on the density profile is greater than that of SIDM0.1, but less noticeable than that of SIDM1 at a given time. Therefore, it appears that if the SIDM model is velocity dependent, then its impact on the density profile, and consequently on the merger rate of PBHs, is more similar to a velocity-independent model with a small value of $\sigma/m$. Note that these models can be diverse, in particular with respect to various constraints, and it would be naive to confidently generalize a result to all SIDM models.

Furthermore, we have investigated the encounter condition of PBHs that may have been randomly distributed in the medium of dark matter halos. Under these assumptions, we have calculated the merger rate of PBHs per halo considering SIDM halo models and compared the results with the corresponding one obtained for CDM halo models. Behaviorally, it has been observed that the merger rate of PBHs in both models is directly proportional to the halo mass. Also, the results indicate that the merger rate of PBHs for SIDM halo models, when $f_{\rm PBH}>0.32$, should be higher than the one extracted from CDM halo models. This means that over time and entering the second epoch (i.e., $\Delta t > t=25 \, t_0$ after the halo virialization), SIDM halo models can potentially amplify PBH mergers compared to the results for CDM halo models. Although we have assumed a strong limit of gravitational focusing for PBH encounters where dissipation by gravitational radiation is more important than dissipation by tidal forces, this condition is not necessarily satisfied. In other words, some processes, such as the disruption of PBH binaries and the torquing effects caused by the surrounding objects, remain as uncertainties in our analysis. In this study, we have focused on PBH binaries formed in dark matter halos in the present-time Universe. However, a large number of PBHs, due to the initial clustering and the chance of proximity, could also have decoupled from the Hubble flow and formed binaries in the early Universe. PBH binaries formed in the early Universe continuously emit gravitational waves, gradually shrink, and eventually merge. However, due to the tidal forces of surrounding PBHs, some PBH binaries may disrupt before those mergers. On the other hand, due to the randomly distributed orbital parameters of PBH binaries in the early Universe, those mergers can happen in the past, present, and future. Consequently, those PBH binaries that are supposed to merge in the present-time Universe will dramatically increase the merger rate today. Also, it has been claimed that in this mechanism of binary formation, for observational relevance, PBHs must constitute only a small fraction of dark matter.

In addition, by considering the P-S halo mass function, we have calculated the merger rate of PBHs per unit volume and per unit time for SIDM halo models and have compared the results with the corresponding findings of CDM halo models. According to the exponential term in the halo mass function, it has been observed that the cumulative merger rate of PBHs decreases with increasing the halo mass in both models. The main reason for such behavior is to reduce the dark matter concentration and increase its velocity dispersion in the larger halos. For this reason, the role of the smallest halos is much more prominent than that of the larger halos in such a way that the merger rate of PBHs depends significantly on the choice of the lower limit of the halo mass. It has also been confirmed that the cumulative merger rate of PBHs for SIDM halo models, while considering $f_{\rm PBH}>0.32$, is higher than the results obtained from CDM halo models. This suggests that the amplification of the merger rate of PBHs also occurs in the cumulative scale of halos in SIDM halo models.

The possibility of binary PBH formation over the age of the Universe due to their random distribution is a good motivation to examine the evolution of the PBH merger rate as a function of redshift. Accordingly, the results indicate that the total merger rate of PBHs is directly related to the redshift in both models. In other words, the PBHs have been more likely to form binaries at higher redshifts than in the present-time Universe. It needs to be highlighted that the PBH merger rate in SIDM halo models, for $f_{\rm PBH}>0.32$, is higher than the results obtained in CDM halo models. This result shows that the SIDM models during the evolution of the late-time Universe have been able to produce more PBH mergers than the CDM models. However, uncertainties in the merger rate of PBHs (e.g., those arising from the shape of the halo mass function) make it difficult to provide a model-independent prediction for the merger rate and the corresponding constraints on the population of PBHs.

Finally, to determine the constraint of PBHs, we have studied the merger rate of PBHs concerning their masses and fraction and have compared them with the mergers estimated by the aLIGO-aVirgo detectors during the third observing run, i.e., $15.3\mbox{-}38.8\,{\rm Gpc^{-3}yr^{-1}}$. The results for CDM halo models are not promising in a way that those do not fall into the aLIGO-aVirgo window for any value of $f_{\rm PBH}\leq1$, whereas the situation is slightly different for SIDM halo models in such a way that the merger rates move closer to the aLIGO-aVirgo window. It is worth mentioning that the related results to SIDM halo models are not strictly valid for $f_{\rm PBH}$ close to unity. That is why we have left this situation out of our findings. Although there are many theoretical uncertainties, it has been observed that within the context of SIDM halo models the merger rate of $10\,M_{\odot}\mbox{-}10\,M_{\odot}$ events can be comparable with the aLIGO-aVirgo window. However, it should be noted that the estimated aLIGO-aVirgo mergers include a range of stellar-mass black holes and are not limited to a specific black hole mass. For instance, the aLIGO-aVirgo detectors recorded four merger events of black holes with a mass of about $10\,M_{\odot}$ during the third observing run, which is proportional to $f_{\rm PBH}\sim\mathcal{O}(10^{-1})$ in SIDM halo models. Hence, within the context of SIDM halo models, one can expect that the potential of dark matter is dominated by a combination of SIDM particles and PBHs.

Given the sensitivity of the aLIGO-aVirgo detectors to probe events up to $z\sim 0.75$ (corresponding to $50~\rm Gpc^{3}$), and in terms of the results obtained for SIDM halo models, it is expected that the aLIGO-aVirgo detectors could record at least ten events annually with $f_{\rm PBH} > 10^{-1}$, while for the detection of at least one event during the same time, the PBH fraction is predicted to be $f_{\rm PBH} > 10^{-2}$. We have also estimated a relation for the fraction of PBHs and their masses, which is consistent with the results obtained in this work. According to the merger rates obtained for different PBH masses, it has been observed that the fraction of PBHs varies inversely with their masses.

In addition to the PBH constraint obtained in this work, other strong observational constraints have been imposed on stellar-mass PBHs that come from the gravitational lensing of type Ia supernovae \cite{Zumalacarregui:2017qqd}, the {\it Planck} data on cosmic microwave background anisotropies \cite{Ali-Haimoud:2016mbv, Serpico:2020ehh}, dynamical processes from star clusters in nearby dwarf galaxies \cite{Brandt, Koushiappas:2017chw}, and the accretion limits from the observed number of x-ray binaries \cite{Inoue:2017csr}.

It should also be noted that the constraints on PBHs is subject to many uncertainties, including different dark matter scenarios (e.g., CDM or SIDM), different conditions that may have been imposed on the structures during collapse and formation, some processes that may lead to the growth (e.g., accretion and merger history) or evaporation (e.g., substantial spin) of PBHs, uncertainties arising from black hole formation scenarios and their contribution to the LIGO-Virgo mergers, and the mass distribution of PBHs. Although the presence of such factors may lead to computational errors, with the development of the instrument and a better understanding of unknown processes, one could attain stronger constraints on PBHs in the future.
%%%%%%%%%%%%%%%%%%%%%%%%%%%%%%%%
\section*{Acknowledgments} 
S.F. and M.F. thank the Research Council of Shahid Beheshti University. The authors gratefully acknowledge the anonymous referees for their constructive comments.
%%%%%%%%%%%%%%%%%%%%%%%%%%%%%%%%

\end{document}